\documentclass[prb,twocolumn,showpacs,preprintnumbers,amsmath,amssymb]{revtex4}
\usepackage{graphicx}
\usepackage{dcolumn}
\usepackage{bm}
\usepackage{trfsigns}
\newcommand \be{\begin{eqnarray}}
\newcommand \ee{\end{eqnarray}}
\newcommand \ba{\begin{align}}
\newcommand {\ket}[1]{|#1\rangle}
\newcommand {\bra}[1]{\langle #1|}
\newcommand {\V}[1]{{\bf #1}}

\newcommand {\p}[1]{\partial_{#1}}
\newcommand {\da}[1]{\mathrel{\overset{\downarrow}{#1}}}
\newcommand {\ov}[1]{\overline{#1}}
\newcommand{\rpm}{\textcircled{$\scriptstyle\pm$}}

\begin{document}
\title{Weyl systems: anomalous transport normally explained}
\author{K. Morawetz$^{1,2}$
}
\affiliation{$^1$M\"unster University of Applied Sciences,
Stegerwaldstrasse 39, 48565 Steinfurt, Germany}
\affiliation{$^2$International Institute of Physics- UFRN,
Campus Universit\'ario Lagoa nova,
59078-970 Natal, Brazil}
\begin{abstract}
The chiral kinetic theory is derived from exact spinor mean field equations without symmetry-breaking terms for large classes of SU(2) systems with spin-orbit coupling. The influence of the Wigner function's off-diagonal elements is worked out. The decoupling of the diagonal elements renormalizes the drift according to Berry connection which is found as an expression of the meanfield, spin-orbit coupling and magnetic field. As special limit, Weyl systems are considered. The anomalous term $\sim\V E\V B$ in the balance of the chiral density appears consequently by an underlying conserving theory. The experimental observations of this term and the anomalous magneto-transport in solid-sate physics usually described by chiral kinetic theory are therefore not a unique signal for mixed axial-gravitational or triangle anomaly and no signal for the breaking of Lorentz-invariance. The source of the anomalous term is by two thirds the divergence of Berry curvature at zero momentum which can be seen as Dirac monopole and by one third the Dirac sea at infinite momentum. During the derivation of the chiral kinetic theory this source by the Dirac sea is transferred exclusively to the Dirac monopole due to the projection of the spinor Wigner functions to the chiral basis. The dynamical result is shown to suppress the anomalous term by two thirds.
\end{abstract}

\maketitle

\section{Introduction}
\subsection{Experimental findings}

Relativistic Fermions with zero mass and consequently linear dispersion have a definite chirality by parallel or anti-parallel spin and motion directions \cite{W29}.   
In condensed matter physics an excitation of chiral mass-less Fermions has been detected in the class of Weyl semi-metals. The Weyl semi-metal with broken time-reversal symmetry is described by such two mass-less Dirac particles with linear dispersion. The chirality is measured here by the photocurrent in response to circularly polarized mid-infrared light \cite{Ma17}. The first Weyl semi-metals have been discovered in $TaAs$ \cite{Xu613,Lu622,Lv15,Hu15,We15} and predicted in \cite{Wa11}. A new type of Weyl semimetal state as been observed in $Mo_xW_{1-x}Te_2$ materials \cite{Bel2016} and in $Ta_3S_2$ as robust Weyl semimetal \cite{Cha16}. 

If the two-band touching points are separated from each other in momentum, the time-reversal symmetry is broken \cite{Sha16}. The band-crossing points acts as magnetic monopoles \cite{D31} being a singular point of Berry curvature and can be described as fictitious magnetic field \cite{SY12a,NN13}. These Weyl points and magnetic monopoles can be fabricated by ultracold atoms and laser-assisted tunneling \cite{DKKS15}. Shubnikov de Haas oscillations \cite{Hu15a} have been detected leading to a Berry phase \cite{B83} accumulation along the cyclotron orbit of $\pi$ indicating Weyl points.
Two linear dispersion bands with touching at four isolated Weyl points in the three-dimensional Brillouin zone have been observed in a double-gyroid photonic crystal \cite{Lu622} where the inversion symmetry breaking is crucial.

Photonic Weyl points have been realized in three-dimensional photonic crystals \cite{Gao16} and the circular photogalvanic effect measure the topological charge of the Weyl nodes which leads to the observation of a quantization of the latter one \cite{Juan17}. Related to this is the topological Chern number which has been used also in studying the chiral magnetic effect in quark systems \cite{Fuk08}.  For a recent review about Weyl and Dirac semi-metals see \cite{AMV18} as well as \cite{Lu2017}.

Though the two chiral populations do not mix without interaction, in parallel electric and magnetic fields charge might flow between Weyl nodes leading to negative magneto-resistance\cite{Xi15}.
This axial current is the chiral anomaly of Adler-Bell-Jackiw \cite{Ad69,BJ69} resulting in the total number of Weyl nodes being even \cite{NN81,NN83}.
This dissipation-less current channel through the vacuum state of a pair of Weyl points causes an enhancement of electric currents. It has been reported a suppression of back scattering in $Cd_3As_2$ and an applied magnetic field lifts this protection leading to large magneto-resistance \cite{Liang14}. Negative longitudinal magneto-resistivity has been observed in $Bi_{0.96}Sb_{0.04}$ \cite{Ki13}, $Na_3Bi$ \cite{Xi15}, $Cd_3As_2$ \cite{Li15},
$TaP$ \cite{Ar16}, $TaAs$ \cite{Hu15a} and $ZrTe_5$ \cite{Li16b} when the magnetic field is parallel to the current \cite{Li16b}. Due to the anomaly an additional current is induced along the magnetic field direction \cite{NN83}. Theoretical this anisotropy in magneto-conductance has been suggested \cite{Aj12,SS13} and
the chiral anomaly might be probed with nonlocal transport since the induced valley imbalance diffuses over long distances \cite{Pa14}. 

In Weyl semi-metals, two opposite Weyl points are connected by arcs in the Fermi contour \cite{De17}. These Fermi arcs lead to unusual quantum interference and has been observed in $TaAs$ and $NbAs$ \cite{Xu613,Lv15a,Ya15,LA15}. Open Fermi arcs lead to unusual magnetic orbits \cite{Po14} and have been suggested to be observable in pyrochlore iridates $Y_2Ir_2O_7$ \cite{Wa11}. The topological protected Fermi arcs are investigated for different minimal models for Weyl semi-metals e.g. in \cite{CKT17} and has been visualized \cite{Ba16}.

Let us shortly remind the main phenomenological ideas. In \cite{Fuk08} the following heuristic discussion has been given. A parallel electric and magnetic field changes the chirality. The Fermi momentum of the right-handed Fermions increases in the electric field
\be
p_F=eEt
\ee
with opposite direction for left-handed ones. The density of left and right-handed Fermions is the product of longitudinal phase-space density $dN_R/dz=p_F/2\pi\hbar$ and the density of Landau levels in traverse direction $d^2N_R/dx dy=eB/2\pi\hbar$ such that the rate of chirality $N_5=N_R-N_L$ is
\be
{d n_5\over d t}={d^4 N_5\over d t d^3x}={e^2\over 2 \pi^2\hbar^2} \V E \cdot \V B.
\label{rate}
\ee
Therefore the term $\V E\V B$ is considered as the origin of non-conservation of chiral charge. However, it can be recasted into a divergence of a quantum current \cite{M19} such that the balance equation is obeyed.

When the chemical potentials of left and right-handed Fermions do not equal, the chiral chemical potential $\mu_5=(\mu_R-\mu_L)/2$ appears besides $\mu=(\mu_R+\mu_L)/2$ and the chiral density is \cite{Fuk08,Li16b}
\be
n_5={\mu_5^3 +\mu_5 (T^2\pi^2+\mu^2)\over 3 \pi^2 v_F^3}.
\label{1}
\ee
The chiral anomaly (\ref{rate}) suggests that the rate is given by
\be
{d n_5\over d t}={e^2\over 2 \pi^2\hbar^2} {\V E}\cdot {\V B}-{n_5\over \tau_v}
\ee
with an assumed scattering time $\tau_v$. Solving in the stationary state 
and $\mu_5\ll \mu,T$ it leads with (\ref{1}) to \cite{Li16b}
\be
\mu_5={3 v_F^3 e^2\over 2 \hbar^2} {\V E\cdot \V B\over \pi^2T^2+\mu^2}\tau_v.
\ee
Since the current density is given by $j=e^2\mu_5 \V B/2\pi^2\hbar^2$ one expects a chiral magneto-conductivity $\sim B^2$ which has been observed \cite{Li16b} and citations therein. Also a linear magneto-conductivity can be as well suggested by a heuristic argument \cite{ZSB12}. There the energy difference $\Delta \epsilon$ of  the Weyl nodes was considered and the associated power cost of the
particle transfer process is $j E=\Delta \epsilon d(N_+ - N_-)/dt$. This provides with (\ref{rate}) a current density
\be
j = {e^2\Delta \epsilon \over 2 \pi^2\hbar^2} B
\ee
derived also by symmetry breaking assumptions \cite{ZSB12}. A suggestion to observe simultaneously chiral magnetic and vorticy effects can be found in \cite{L11,L14} where the vanishing of the chiral magnetic current in the bulk was reported. 

However, there are some problems to link the observations really to chiral anomaly. Recently it was demonstrated that longitudinal negative magnetoresistance can appear also in conventional centrosymmetric and time-reversal invariant conductors \cite{AS18}. Consequently it is not a unique signal for chiral anomaly.
Extremely large magneto-resistance $[\rho(h)-\rho(0)]/\rho(0)$ has been observed \cite{Sh15} in $NbP$ being five times larger than in $WTe_2$ \cite{Ali14} and twice as large as in $TaAs$ \cite{Zh17}. In $TaP$ it was reported \cite{Ar16} that the negative magneto-resistance cannot be linked to the chiral anomaly when the Fermi surface connects both Weyl nodes. In \cite{KWSSSNBUWFY16} it is predicted that chiral anomaly can be realized in $NbP$ if the Fermi energy is driven to the Weyl points by electron doping.
This is in contrast to the observation of no negative magnetic conductance in $NbP$ reported in \cite{SKNDP17}. A crossover from nearly parabolic behavior at
low fields to linear behavior at high fields is seen which suggests
different scattering mechanisms in a disordered environment. In \cite{KRP17} a breakdown of the chiral anomaly in Weyl semi-metals was reported in a strong magnetic field since a sizable energy gap opens
up due to the mixing of the zeroth Landau levels associated with the opposite-chirality Weyl points.

Since controversial results in the literature are attributed to current jetting effects by the geometry of voltage and current contacts on the sample which is not related to chiral anomaly \cite{Ar16}.
The Berry phase calculated from the frequency of Shubnikov de Haas oscillations shows no additional phase factor and no evidence for longitudinal
magneto-resistance which seems to rule out chiral anomaly in $NbP$. Combined Hall and magneto-transport data suggest that the linear magneto-resistance observed in $NbP$ is probably due to charge carrier mobility fluctuations.

Having observed some doubts in the common conviction that chiral anomaly is observed in solid state physics, let us discuss an even more puzzling statement that Lorentz-symmetry breaking is believed to be observed.

\subsection{Chiral anomaly and gravitational anomaly}

The Lorentz symmetry is the basics of physical laws to be independent of the frame. The search of Lorentz-symmetry violation is getting a new drive with the discovery of pairs of particles with tilted Weyl cones such that they have both positive dispersions called Weyl semi-metals of type II \cite{ASGWWTDB15,Xu17}. The absorption of circular polarized light in various tilted Dirac cones has been treated in \cite{MC17}. Experimental evidence of vanishing magneto-thermoelectric conductance in $NbP$ seems to support the existence of such axial-gravitational anomalous terms \cite{gooth17}.

Lorentz invariance leads to three types
of Fermions: Dirac, Weyl and Majorana. A special choice of Dirac Matrices (Majorana) can render the Dirac equation real and $\Psi=\Psi^*$ represents Majorana Fermions. Helicity or handedness is twice
the value of the spin component of a particle along the direction of its
momentum. This helicity is frame-dependent for massive particles. The left and right-handed projection can be realized by the matrices $(1\mp\gamma_5)/2$ with $\gamma_5=i\gamma^0\gamma^1\gamma^2\gamma^3$ which anticommute with all Dirac matrices. This chiral projections can be made Lorentz-covariant \cite{CPWW13}. However, chirality is not conserved even for a free
particle because $\gamma_5$ does not commute with the mass term in
the Dirac Hamiltonian.
If the Fermions are massless, the Dirac equation decouple into a left- and right-handed one and the chirality and helicity coincides and we have Weyl fields \cite{Pa11} with a possible triply degenerate point \cite{Lv17}. 

Very often explicitly a charge anomaly as axial chemical potential term in the relativistic Lagrangian is introduced \cite{ZB12,CPWW13,Bu15,Bu16,JHK17} ad hoc
\be
{e^2\over 16 \pi^2\hbar^2 c}(\V b\V r-b_0 t)\varepsilon^{\mu\nu\alpha\beta}F_{\mu\nu}F_{\alpha\beta}
\label{axion}
\ee
with chiral gauge fields $(b_0,\V b)$ also called axion fields \cite{GT13}. The electrodynamics assuming explicitly such a chiral breaking term has been treated in \cite{QCH17}. This gauge field is added as effective theta angle to the Lagrangian \cite{Fuk08}. 
This yields the non-conservation of the axial current $J_5=\ov {\Psi} \gamma^\mu \gamma^5\Psi$ in the form
\be
\partial_\mu J_5^\mu= 2i m \ov \Psi \gamma^5\Psi + {e^2\over 16 \pi^2\hbar^2 c}\varepsilon^{\mu\nu\alpha\beta}F_{\mu\nu}F_{\alpha\beta}
\label{4}
\ee
and has been shown to lead to negative quadratic longitudinal magneto-resistance \cite{Bu15}. The first term of (\ref{4}) vanishes for zero mass leading to conservation of zero-mass Dirac particles while the second part expressed by the field tensor $F_{\mu \nu}$ violates axial current conservation. 

Due to the axial non-conservation it is called also mixed axial-gravitational anomaly and claimed to violate Lorentz symmetry \cite{ZB12,JHK17,Xu17,gooth17}. In \cite{L14} it was pointed out that the divergence of covariant currents is uniquely defined, while the divergence of consistent currents depends on
specific regularization schemes which freedom allows the definition of an
exactly conserved electric current. It should lead to the physical observation that charges separate at the edges perpendicular to the magnetic field
even when there is no bulk current according to kinetic theory.

All these Lorentz-symmetry breaking approaches rely on the axial coupling of the field (\ref{axion}). In \cite{DelCima2017} it has been shown recently that a proper subtraction scheme of the infrared divergences shows that such terms do not appear. Therefore the claim of Lorentz-symmetry violation and consequently gravitational anomaly is not well founded theoretically. It is even not impossible that the anomalous term (\ref{rate}) appears also by approximating of an underlying symmetry-respecting theory. 
Here we will show exactly this path how from a conserving theory a seemingly symmetry-breaking transport theory appears.

In fact the Lorentz-invariant chiral kinetic theory can be derived from the quantum kinetic approach \cite{GLPWW12,CPWW13,MT14,GPW17,HPY17}. In \cite{GPW17} the three dimensional chiral kinetic equation is derived from four dimensional Wigner function in powers of space-time derivatives. It is found that the kinetic equation is not uniquely determined and one needs to build up such chiral kinetic theory directly from a covariant Wigner equation. Collisional processes create side jumps and their frame dependence in the parametrization of the Wigner functions is discussed in \cite{HPY17} where the explicit demand of Lorentz covariance leads to modified side jumps. 
The different choices of frames are discussed in the derivation of \cite{HSJLZ18}.

\subsection{Summary of problems and intention of the present paper}

The basis of the experimental interpretation \cite{Hu15,Xu17,gooth17} of having observed chiral anomaly and breaking of conservation laws like mixed axial-gravitational anomaly is the $\V E\V B$ term in the chiral density balance (\ref{rate}). This term has been suggested by anomalous terms in the field-theoretical Lagrangian coming from triangular anomaly \cite{J12,BKY14,L16} known as Adler-Jackew-Bell anomaly \cite{Ad69,BJ69,NN83}. Since the path from this symmetry-violating assumptions to the final non-conservation form is well worked out \cite{ZB12a}, it leads to the impression as if the opposite conclusion would be unique as well and it would be a unique signal of symmetry breaking. However, this is not a one-to-one correspondence. One cannot conclude backwards from the observed term to a symmetry-breaking field-theoretical assumption. Why should there not be another path to obtain the same term (\ref{rate}) from a conserving theory without the described field-theoretical assumptions? In fact the paper here will show such a possible way. Therefore we cannot interpret the $\V E\V B$ term observed in solid state physics as a unique sign of Adler-Jackew-Bell anomaly \cite{DJ74,NN83,BKY14}.

The second problem concerns the interpretation of the term (\ref{rate}) arising by magnetic monopoles \cite{D31,H74} and the divergence of the Berry-curvature at zero momentum \cite{Fang03}. We will see that a part of this anomalous term comes actually from the Dirac sea at infinity momentum and not exclusively from the zero momentum which would be a Dirac monopole.

Here in this paper we will derive the anomalous chiral transport from a conserving Hamiltonian without symmetry breaking and without the help of anomalous terms. We use the non-relativistic formulation with a proper spin-orbit coupling to show this for large classes of spin systems. The chiral transport for Weyl systems is then obtained by the infinite-mass limit where only the chosen spin-orbit coupling describing chiral particles remains. 

The paper is ordered as follows. First, in the next chapter we repeat shortly the chiral kinetic theory based on the phenomenological completion of Hamilton equations by Berry curvature as one finds in the literature. This establishes the basis on which most experimental results are explained. In chapter III we will summarize the exact coupled quantum-spin quasiclassical kinetic equation on the mean-field level. In chapter IV this spinor equation is then transformed into the heuristic one of chiral kinetic theory with specification of all necessary approximations. This will represent the main work of the paper. In chapter V we come back to the chiral anomaly and compare the results arising once from chiral kinetic theory and once from the exact quasiclassical kinetic equation. In chapter VI we summarize and conclude. 

\section{Chiral kinetic theory}
Many theoretical approaches link the anomalous magnetic conductance to the chiral anomaly. Usually \cite{KKS14,Zy17,Be17} the Berry curvature $\V \Omega_\pm=\partial_k\times \V a_\pm=\pm \hbar {\V k}/2 k^3$ modification of the Boltzmann equation is used. Here the Berry connection \cite{B83} $\V a_\pm=i\hbar \bra{\pm}\partial_k\ket{\pm}$ is the measure of the overlap of wave functions \cite{Fang03}. It is assumed that the Hamilton equations of quasiparticle trajectories become modified due to this Berry curvature as \cite{SY12a,SS13,KKSL14,KKS14,Zy17}
\be
\dot {\V r}=\V v+\dot {\V k}\times \V \Omega_\pm,\qquad \dot {\V k}=e\V E+\dot {\V r}\times e\V B
\label{haldane}
\ee
which makes the equations symmetric \cite{Ha93}. 
Eq. (\ref{haldane}) can be disentangled to yield
\be
\dot {\V r}&=&{\V v+e\V E\times \V \Omega_\pm+e \V B\V v\cdot \V \Omega_\pm \over 1+e\V B\cdot \V  \Omega_\pm}\nonumber\\
\dot {\V k}&=&{e\V E +\V v\times e\V B+e^2\V \Omega_\pm \V E\cdot \V B\over 1+e\V B\cdot \V  \Omega_\pm}.
\label{rp}
\ee
This determines the drift of the phenomenological kinetic equation
\be
\dot f_\pm+\dot {\V r}\, \partial_{\V r}\, f_\pm+\dot {\V k}\,\partial_{\V k} \,f_\pm=I_\pm
\label {phenom}
\ee
for the distribution of both chiral particles $f_\pm$ with some collision integrals $I_\pm$.
The anomalous velocity term in the equations of motions has been treated with Bloch electrons \cite{Ha93} where the anomalous Hall effect has been shown as a Fermi-liquid property.

Many experimental facts are derived from this chiral kinetic equation (\ref{phenom}).
In \cite{Zy17} a magneto-conductivity of 3/2 power of the magnetic field is found and the time-reversal symmetry breaking results in linear form.
Magnetic and anomalous contributions to the Nernst coefficient has been computed as the transverse electrical response to a longitudinal thermal gradient in absence of a charge current \cite{Sha16}. A violation of the Wiedemann-Franz law has been found within this phenomenological Boltzmann approach \cite{Kim14}. The thermoelectric properties in Weyl and Dirac semi-metals are investigated in \cite{LPG14}. Assuming a Lorentz-invariance breaking term which tilts the Weyl cone, the anomalous Nernst and Hall effect have been reported in \cite{FZB17}.

A more refined kinetic theory by diagonalizing the Heisenberg equation was 
given in \cite{SCM17} where still the term $\mu_5\sigma_z$ is added to the Hamiltonian. The derivation from the density-matrix equation is presented in \cite{WT11}. Multi-Weyl semi-metals have been treated in \cite{PWMM17}. The Weyl semi-metals with spin-orbit coupled impurities are described with matrix Green's functions in \cite{LHC17} and an anomalous Hall effect in pyrochlore iridates was predicted \cite{YJR11}. A systematic derivation of chiral kinetic theory is presented in \cite{MT14} and extended to chiral relativistic plasmas \cite{MT14a}. The field theoretical worldline construction of a covariant chiral kinetic theory can be found in \cite{MV17} and a gauge-invariant formalism with Berry curvature is presented in \cite{Be17}. The connection between magnetic response of strongly interacting matter and axial anomaly has been worked out in \cite{NS06}.

Alternatively, the Berry curvature can be incorporated by a fictitious vector potential which leads to nontrivial commutator relations of Poisson brackets \cite{SY12,SS13}
\be
\{p_i,p_j\}&=&-{\epsilon_{ijk}eB_k\hbar \over 1+e\V B\cdot \V \Omega_\pm},
\quad
\{x_i,x_j\}={\epsilon_{ijk}\Omega_{k\pm}\hbar \over 1+e\V B\cdot \V \Omega_\pm}
\nonumber\\
\{p_i,x_j\}&=&\hbar {\delta_{ij}+\Omega_{i\pm} eB_j\over 1+e\V B\cdot \V \Omega_\pm}
\ee
with the invariant phase space
\be
d\Gamma=(1+e \V B \cdot \V \Omega_\pm) {d\V k d \V x  \over (2\pi)^3}.
\label{G}
\ee
In this way the phenomenological kinetic equation (\ref{phenom}) is reproduced \cite{NN13}.
The same equation of motions have been derived fixing the non-Abelian U(2)-gauge freedom  by diagonalizing the Hamiltonian and neglecting certain off-diagonal elements \cite{SY12a}. A relativistic derivation of (\ref{phenom}) was presented from the Dirac equation in \cite{MT14}. A manifest Lorentz-covariant equation (\ref{phenom}) was derived \cite{CPWW13} assuming a Berry connection term as a vector potential in momentum space  in the action. Compared to (\ref{rp}) it possesses additional terms coming from 4-vector derivatives.  

The phenomenological chiral equation (\ref{phenom}) results into the charge (non-conserving) balance \cite{SY12,SS13,NN13,KKS14,MT14a}
\be
\dot n_\pm+\nabla \V j_\pm=-{\xi e^2\over 4 \pi^2\hbar^2} \V E\cdot \V B
\label{balancerate}
\ee
with the particle number current
\ba
\V j_\pm\!=&\!-\!\int {d^3 k\over (2\pi\hbar)^3} \epsilon\left [ \partial_{\V k} f\!+\!\left (\V \Omega_\pm\cdot \partial_{\V k} f\right )e \V B\!+\!\V \Omega_\pm \times \partial_{\V r} f \right ]
\nonumber\\
&+e\V E\times \int {d k\over (2\pi)^3} \V \Omega_\pm f.
\end{align}
and since $\p {\V k}\V \Omega_\pm=0$ we have with Gau\ss{} integral theorem
\be
\xi=\int {d^3 k\over 2\pi}\V \Omega_\pm \p {\V k} f={1\over 2 \pi} \left . \int d\V A\cdot \V \Omega_\pm f\right |_{k=0}^{k=\infty}=\pm 1
\label{topo}
\ee
which is the monopole charge inside the Fermi surface \cite{SY12,D31} at zero temperature.
Therefore the balance equation (\ref{balancerate}) clearly shows the chiral anomaly (\ref{rate}).

\section{Quantum kinetic equation}
Now we will present the starting kinetic theory of coupled Wigner functions from which we will derive the chiral kinetic equation. Let us here outline shortly the exact quantum kinetic equation of a mean-field Hamiltonian with SU(2) structure and electromagnetic fields \cite{M15}. In this way we will show that large classes of spin-orbit coupled systems as well as Weyl systems and graphene illustrated in table~\ref{tab1} are possible to recast into a form of chiral kinetic theory when performing certain approximations.

\subsection{Coupled spinor equations}\label{exact}
 
The effective Hamiltonian with Fourier transform of spatial coordinates $\V r\to \V q$ reads
\be
H=\epsilon_k+\Sigma_0(\V k,\V q,t)+\V \sigma \cdot \V \Sigma (\V k,\V q,t)
\label{ham}
\ee
with the single-particle energy $\epsilon_k$ and the Pauli matrices $\V \sigma$. The scalar selfenergy consists of the electrostatic potential $\Sigma_0=\Sigma_o^{\rm MF}+e \Phi(\V q, t)$ and the scalar meanfield
\be
&&\Sigma^{\rm MF}_0=
{n} V_0+{\V s}\cdot \V V. 
\ee
Here the particle density ${n}=\sum\limits_k  f$ is given by the scalar Wigner function $f$ and the spin- polarization ${\V s}=\sum\limits_k  \V g$ is given by the vector Wigner function $\V g$. Both appear as frequency integral, $\int {d\omega \over 2 \pi } G^<=\rho=f+\V \sigma \cdot \V g$, over the correlation (Green's) function \cite{M17b} $G^<_{12}=\langle a^+_1 a_2\rangle$.
We can think these meanfields as arising from scalar impurity $V_0(q)$ or magnetic impurity $\sigma\cdot \V V(q)$ scattering.

The vector self energy turns out to possess three parts
\be
\V {\Sigma}={\V \Sigma}^{\rm MF}(\V p,\V q,t)+\V b(\V p)+\mu_B \V B.
\label{Sv}
\ee 
Besides the Zeeman-term of magnetic field, $\mu_B \V B$, it has the vector meanfield
\be
\V \Sigma^{\rm MF}=
{\V s}\, V_0+{n} \V V
\ee
and the spin-orbit coupling vector
\be
\V b \cdot \V \sigma=A(\V k) \sigma_x -B(\V k) \sigma_y +C(\V k) \sigma_z.
\label{so}
\ee
With different choices of this vector we can describe the mean-field dynamics of great classes of systems as illustrated in table~\ref{tab1}. The idea is to realize graphene and Weyl Hamiltonians by the infinite-mass limit \cite{M16} which kills the quasiparticle energy $\epsilon_k$ and leaves only the spin-orbit coupling of exactly the form of chiral Hamiltonian. 

\begin{table}
\caption{\label{tab1} Selected 2D and 3D systems described by (\ref{so}) taken partially from \cite{Chen09,cserti06}. The * denotes the additional infinity mass limit of nonrelativistic particle to generate chiral dispersions.}
\begin{align*}
&
\begin{array}{llll}
{\rm 2D-system} & A(k) & B(k) & C(k)\cr
\hline
{\rm Rashba} & \beta_R k_y & \beta_R k_x &\cr
{\rm Dresselhaus} [001] & \beta_D k_x & \beta_D k_y &\cr
{\rm Dresselhaus} [110] & \beta k_x & -\beta k_x &\cr
{\rm Rashba-Dresselhaus} & \beta_R k_y-\beta_D k_x & \beta_R k_x-\beta_D k_y &\cr
{\rm cubic}\, {\rm Rashba (hole)} & i{\beta_R \over 2}(k_-^3-k_+^3)&{\beta_R \over 2}(k_-^3+k_+^3) &\cr
{\rm cubic}\, {\rm Dresselhaus} & \beta_D k_x k_y^2& \beta_D k_y k_x^2 &\cr
{\rm Wurtzite}\, {\rm type} & (\alpha +\beta k^2 ) k_y & (\alpha+\beta k^2 ) k_x&\cr
*{\rm single-layer}\, {\rm graphene} & v k_x & -v k_y&\cr
*{\rm bilayer}\, {\rm graphene} & {k_-^2+k_+^2\over 4m_e} & {k_-^2-k_+^2\over 4m_e i}& 
\end{array}
\end{align*}


\begin{align*}
&
\begin{array}{llll}
{\rm 3D-system} & A(k) & B(k)& C(k) \cr
\hline
{\rm bulk}\, {\rm Dresselhaus} & k_x(k_y^2-k_z^2) & k_y(k_x^2-k_z^2) &k_z(k_x^2-k_y^2) \cr
{\rm Cooper pairs} & \Delta & 0 & {p^2\over 2 m} -\epsilon_F \cr
{\rm extrinsic}\, {\rm spin-orbit} &&&\cr
 \beta={i\over \hbar}\lambda^2 V(k) &q_y k_z-q_z k_y&q_z
k_x-q_x k_z &q_x k_y-q_y k_x \cr
{\rm neutrons \, in\,  nuclei} &&&\cr
 \beta={i} W_0 (n_n+{n_p\over 2}) &q_z k_y-q_y k_z&q_x k_z -q_z
k_x&q_y k_x -q_x k_y\cr
*{\rm Weyl}\,\,{\rm materials} &v k_x &-v k_y &v k_z
\end{array}
\end{align*}
\end{table}

The coupled kinetic equations of scalar and vector Wigner functions read \cite{M15}
\be 
(\p t+{ {\cal F}}\V {\p {\V k}}+\V { v}\V {\p {\V r}}) f
+\V A\cdot \V g &=&0\nonumber\\
(\p t+{ {\cal F}}\V {\p {\V k}}+\V {v}\V {\p {\V r}}) \V g\,
+\,\V A \, \, f
&=&{ \!2 (\V {\Sigma}\times   \V g)}.
\label{kin}
\ee
The coupling between the two equations is caused by the vector selfenergy (\ref{Sv})
\be
A_i = \V {\p {\V k}}{\Sigma_i}\V {\p {\V r}}-\V {\p {\V r}}{\Sigma_i}\V {\p {\V k}}+(\V {\p {\V k}}{\Sigma_i} \times e\V B)\V {\p {\V k}}
\label{A}
\ee 
 with the effective velocity and effective Lorentz force
\be
\V {v}={\partial \epsilon_k\over \partial k}+\V {\p {\V k}} \Sigma_0,\qquad
{{\cal F}}=(e \V E - \V {\p {\V r}}
\Sigma_0+e \V {v} \times  \V B ).
\label{lorentz}
\ee
The right-hand side of (\ref{kin}) shows the spin precession term which is the reason for the anomalous Hall effect \cite{M15}. In principle we could add the collision integrals on the right-hand side.

The stationary solution shows a two-band splitting \cite{M15}
\be 
 \rho({ \varepsilon})=\sum\limits_{\pm} P_\pm f_\pm
&=& {f_++f_-\over 2}+\V \sigma\cdot {\V e} \,\,{  f_+-f_-\over 2}
\nonumber\\
&=&f_{\rm eq}+\V \sigma \cdot \V e \, g_{\rm eq}
\label{solF}
\ee
with the Wigner functions $f_\pm=f_0(\epsilon_k\pm|\V \Sigma|)$,
the selfconsistent meanfield $\epsilon_k(r)={k^2\over 2 m}+\Sigma_0(k,r)$, and
a selfconsistent precession \cite{M15}
${\V e(p,r)=\V \Sigma/|\V \Sigma|}$. The projectors can be defined as $ P_\pm=\frac 1 2 (1\pm \V e\cdot \V \sigma)$.

The equation system (\ref{kin}) is a rewriting of Green's functions obeying carefully the noncommutativety of spinors and gauge-invariance up to second order space-time derivatives. The used gradient approximation \cite{M15} to derive (\ref{kin}) affects not the spinor structure since the commutators have been considered exactly. To extend (\ref{kin}) from the quasiclassical form to the quantum form, one would merely replace the Poisson bracket with respect to $\V r,\V k$ in (\ref{kin}) by commutators allowing to describe the quantum Hall effect \cite{M15}. With this restriction we call (\ref{kin}) the exact mean-field or spinor kinetic equations. They represent the coupled scalar and vector equations of the Wigner function formalism \cite{VaGy87} but extended here to meanfields and spin-orbit coupling.

Though it has been shown that (\ref{kin}) provide anomalous Hall conductivity \cite{M15}, spin-density waves \cite{M15a} and graphene transport \cite{M16} directly, it is now useful to map this equation system to the helicity basis used in the literature. This will enlighten the connection to anomalous transport with Berry curvatures. We will perform the derivation for the general Hamiltonian (\ref{ham}) such that large classes of systems are covered as illustrated in table~\ref{tab1}. In the end we will use the infinite-mass limit of non-relativistic particles \cite{M16} to produce the chiral dispersion according to table~\ref{tab1}.

\section{Transformation to chiral kinetic equation}\label{transform}
\subsection{Helicity basis}

For this purpose we go into the helicity basis of (\ref{ham})  which means we use the eigenstates 
\be
H\ket \pm =\epsilon^\pm \ket \pm
\ee
and using the notation $\Sigma=|\V \Sigma|$ and $\Gamma=\sqrt{\Sigma^2-\Sigma_z^2}$
\be
\Sigma_x+i \Sigma_y=\Gamma {\rm e}^{i\varphi}
\label{polar}
\ee
we have
\be
\epsilon^\pm&=&\epsilon\pm \Sigma\nonumber\\
\ket \pm&=&{1\over \sqrt{2}}\begin{pmatrix}-{\rm e}^{-i \varphi}\sqrt{1\pm {\Sigma_z\over \Sigma}}\cr \mp \sqrt{1 \mp {\Sigma_z\over \Sigma}}\end{pmatrix}.
\label{pm}
\ee
The Hamiltonian (\ref{ham}) becomes
diagonal
\be
\bar H=U^+HU=\begin{pmatrix}\epsilon^+&0\cr 0&\epsilon^-\end{pmatrix}
\label{hbar}
\ee
by the transformation matrix $U=(\ket +,\ket -)$ which means we have the property
\be
\overline{\V \sigma\cdot \V \Sigma}=U^+ (\V \sigma\cdot \V \Sigma) U=\Sigma \sigma_z.
\label{prop1}
\ee

Since the transformed spin-projection operators read
\be
\bar P_+=\begin{pmatrix}1&0\cr 0&0\end{pmatrix},\quad \bar
P_-=\begin{pmatrix}0&0\cr 0&1\end{pmatrix}
\ee
the equilibrium Wigner function (\ref{solF}) becomes diagonal
\be
 {\bar \rho}_{\rm eq}=\sum\limits_{i=\pm} P_i f_i=\begin{pmatrix}f_+&0\cr 0&f_-\end{pmatrix}.
\ee

In nonequilibrium this situation is much more complicated since we have to take into account the momentum and time dependence of the transformation matrix. As a result we will see that the transformed Wigner functions will maintain 4 components. Indeed we have
\be
 \rho=f+\V \sigma\cdot \V g=\begin{pmatrix}
f+g_z &g_x-i g_y\cr g_x+i g_y&f-g_z
\end{pmatrix}
\ee
and the transformed equations read
\ba
\bar \rho =U^+ \rho U=\begin{pmatrix}
f_{+\!+} &f_{+\!-}\cr f_{-\!+}&f_{-\!-}
\end{pmatrix}=f+\sigma_z g +i\sigma_y \Delta +\sigma_x f_3.
\label{hel}
\end{align}
Here we have introduced besides the two Wigner functions,
\be
f=\frac 1 2 (f_{+\!+}+f_{-\!-});\qquad g=\frac 1 2 (f_{+\!+}-f_{-\!-}),
\ee
 which take the same form in equilibrium as our untransformed spinor ones,
also the off-diagonal Wigner functions
\be
f_3=\frac 1 2 (f_{+\!-}+f_{-\!+});\qquad \Delta=\frac 1 2 (f_{+\!-}-f_{-\!+}).
\ee

Now we are going to transform the equation system (\ref{kin}). Therefore we multiply the second equation of (\ref{kin}) with $\V \sigma$ and add both equations.
With the help of the identity 
\ba
&\V c\cdot \V g+ (\V \sigma \cdot \V c) f-2 \V \sigma \cdot (\V \sigma\times \V g)=
\nonumber\\&
\V \sigma \cdot {\V c+2 i\V \sigma\over 2}  \rho+ \rho \V \sigma \cdot {\V c-2 i\V
  \sigma\over 2}
\label{identity}
\end{align}
we obtain an operator equation for $ \rho=f+\V \sigma \cdot \V g$
\be
D^t \rho +\frac 1 2 \{\V\sigma\cdot \V A,\da \rho\}+i[\V \sigma\cdot \V \Sigma,\rho]=0
\label{kin1}
\ee
where the $\{\}$ brackets denote the anticommutator and $[]$ the commutator. We have abbreviated the differential operator
\be
D^t=\p t+\V v \cdot \p {\V r}+ {\cal F} \cdot \p {\V k}
\label{D}
\ee
with the velocity and force (\ref{lorentz}). Since $\vec A$ of (\ref{A}) is a differential operator we will specify on which object it acts by a $\downarrow$ above the variable.  

We transform now the equation (\ref{kin1}). The first term becomes
\be
U^+D^t \da \rho U&=&D^t\bar \rho -D^t\da {U^+}\rho U-U^+\rho D^t \da U
\nonumber\\
&=&D^t \bar \rho+[U^+D^t U,\bar \rho]
\ee
where we have used the unitary property $U^+U=1$ and consequently $(\partial U^+)U=-U^+\partial (U)$. The second term is treated along the same line
\ba
U^+\{\V\sigma\cdot \V A,\da \rho\}U&=\{\overline{\V \sigma\cdot \V A},\da {\bar \rho}-\da {U^+}U {\bar \rho}-{U^+}\da U {\bar \rho}\}
\nonumber\\
&=\{\overline{\V \sigma\cdot \V A},\da {\bar \rho}\}+\{\overline{\V \sigma\cdot \V A},[U^+\da U,\bar \rho]\}.
\end{align}
The last term is simply
\be
U^+[\V\sigma\cdot \Sigma,\rho]U=[\overline{\V \sigma\cdot \V \Sigma},\bar \rho]=[\Sigma\sigma_z,\bar \rho]
\ee
where we used (\ref{prop1}).

All terms together one sees that the kinetic equation (\ref{kin}) translates after transformation into an equation for the chiral Wigner functions (\ref{hel})
\be
D^t\bar \rho&+&[U^+D^t U,\bar \rho]+\frac 1 2 \{\overline{\V \sigma\cdot \V A},\da {\bar \rho}\}+\frac 1 2 \{\overline{\V \sigma\cdot \V A},[U^+\da U,\bar \rho]\}
\nonumber\\&+&i [\Sigma\sigma_z,\bar \rho]=0.
\label{kin2}
\ee

Further progress is made if we explicitly calculate the appearing derivative matrix $U^+\partial U$. With the help of 
(\ref{pm}) one has explicitly the off-diagonal and diagonal Berry connections
\be
\V b_\mp=i \bra \mp \V \partial \ket { \pm}&=&{\V \partial
\varphi\over 2}{\Gamma\over \Sigma} 
\pm
{i\Sigma\over 2 \Gamma} \V \partial \left ({\Sigma_z\over \Sigma}\right )
\nonumber\\
\V a_\pm=i \bra \pm \V \partial\ket \pm&=&{1\over 2}\left (1 \pm {\Sigma_z\over \Sigma}\right ) \V \partial \varphi
\label{derivh}
\ee
with $\Gamma=\sqrt{\Sigma^2-\Sigma_z^2}$. In principle any variable like space, time, and momentum can serve as derivative $\partial$. We will restrict later exclusively on momentum derivatives $\partial =\hbar \p {\V k}$.
The Berry curvature is given by the curl of the band-diagonal Berry connection
\ba
&\pm \V \Omega =\V {\partial} \times \V a_\pm=i\bra {\V {\partial} \pm}\times \ket {\V {\partial} \pm}
\nonumber\\
&\!=\!\pm {1\over 2 \Sigma^3}\left (\Sigma_x \V {\partial} \Sigma_y \!\times\! \V {\partial} \Sigma_z\!+\!\Sigma_z \V {\partial} \Sigma_x \!\times\! \V {\partial} \Sigma_y\!+\!\Sigma_y \V {\partial} \Sigma_z \!\times\! \V {\partial} \Sigma_x \right ).
\end{align}
We want to note that this expression of the Berry connection is a generalization of the known forms since it includes besides the magnetic field also the mean field as well as spin-orbit coupling according to (\ref{Sv}).

Abbreviating $\V b=\V b_\pm$ and introducing conveniently 
\be
\V a={\Sigma \mp \Sigma_z\over \Gamma }\V a_\pm
\ee
we can express
\be
\V \partial \varphi={2 \Sigma \over \Gamma} \V a;\qquad \V \partial \left ({\Sigma_z\over \Sigma}\right )=\pm 2 i{\Gamma\over \Sigma} (\V b-\V a).
\ee 
It is easy to see that 
\be
\V a\times \V b&=&\mp {i\over 2} \V \Omega
\nonumber\\
\V \partial \times \V a&=&\V \partial \times \V b= {\Sigma_z\over \Gamma} \V \Omega.
\label{rel1}
\ee

The occurring derivative transformation matrix takes the form
\be
U^+\V \partial U=-i \V a \left ({\Sigma \over \Gamma} I+\sigma_x+{\Sigma_z\over \Gamma}\sigma_z \right )\pm (\V b-\V a)\sigma_y.
\label{UdU}
\ee

\subsection{Transformation of kinetic equation into helicity components}
Now we have all expressions in terms of Pauli matrices which makes it possible to use the commutator and anticommutator property
\be
[\sigma_i,\sigma_j]=2 i\epsilon_{ijk}\sigma_k;\qquad \{\sigma_i,\sigma_j\}=2 \delta_{ij}.
\ee
This allows us to work out (\ref{kin2}) together with (\ref{hel}) to obtain the kinetic equations for the diagonal and off-diagonal elements. We abbreviate besides the total drift (\ref{D}) also the partial drift
\be
D_c=\V {\p {\V k}}{c}\V {\p {\V r}}-\V {\p {\V r}}{c}\V {\p {\V k}}+(\V {\p {\V k}}{c} \times e\V B)\V {\p {\V k}}
\ee
such that the coupling term (\ref{A}) might be written $\V A=D_{\V \Sigma}$.
Since the Berry-curvature terms $\V a$ and $\V b$ appear due to the corresponding derivatives, we understand 
\be
\tilde D_{\V a}&=&\V {a_k}\V {\p {\V r}}-\V {a}_r\V {\p {\V k}}+(\V {a}_k \times e\V B)\V {\p {\V k}}
\nonumber\\
{\V a}_D&=&\V a_t+\V v \V {a_r}+{\cal F} \V {a_k}.
\label{deriv}
\ee
Here we have in principle the possibility to consider the Berry curvature in space, $\V a_r$, time $\V a_t$, and momentum $\V a_k$. However for legibility we will only concentrate on the momentum curvature $\V a=\V a_k$. In fact, one can see later that the time-dependence lead to higher-order derivatives.

Since $\V A$ is a differential operator we calculate the commutators with the help of (\ref{UdU})
\be
\overline{\V \sigma\cdot \V A}&=&U^+D_{\V \sigma\cdot \V \Sigma} U=D_{\overline{\V \sigma\cdot \V \Sigma}}+[U^+D_U,\overline{\V \sigma\cdot \V \Sigma}]
\nonumber\\&=&D_\Sigma \sigma_z-2 \Sigma \tilde D_{\V a}\sigma_y\pm 2 i \Sigma \tilde D_{\V b-\V a} \sigma_x.
\ee
Further, one obtains the following intermediate steps
\be
\frac 1 2 \{\overline{\V \sigma\cdot \V A},\da {\bar \rho}\}&=&
D_\Sigma f \sigma_z+D_\Sigma g-2 \Sigma \tilde D_{\V a} f\sigma_y-2 i \Sigma \tilde D_{\V a} \Delta \nonumber\\
&&\pm 2 i \tilde \Sigma D_{\V b-\V a} f \sigma_x \pm 2 i \Sigma \tilde D_{\V b-\V a} f_3.
\ee
The derivative of the transformation matrix in (\ref{kin2}) reads
\be
[U^+\partial U,\bar \rho]&=&\!-\!2 \V a \left (g\!-\!{\Sigma_z\over \Gamma} f_3\right )\sigma_y
\!+\!2 i [\V a \Delta\mp (\V b\!-\!\V a) f_3] \sigma_z 
\nonumber\\&&+2 i \left [-\V a{\Sigma_z \over \Gamma} \Delta\pm (\V b-\V a) g\right ] \sigma_x
\ee
where the Berry curvatures correspond to the derivatives $\partial$.
This allows to calculate
\ba
\frac 1 2 \!\left \{
\overline{\V \sigma\cdot \V A},\!\left [U^+\da U,\bar \rho\right ]
\!\right \}
&=-2 i \Delta \tilde D_{\V a}\Sigma  \pm 4\Delta {\Sigma\Sigma_z\over \Gamma}(\V a\times \V b) \cdot e\V B 
\nonumber\\
&\pm f_3\, 2 i \tilde D_{\V b\!-\!\V a}\Sigma.
\end{align}
Finally we have
\be
i [\Sigma\sigma_z,\bar \rho]=-2 \Sigma f_3 \sigma_y+2 i\Sigma \Delta \sigma_x.
\ee

Now we can write the transformed kinetic equation (\ref{kin2}) together. This will lead to the two diagonal equations $(f+ g \sigma_z)$ which we write $(f\rpm g)$ to distinguish from the Berry curvature $\pm$. We obtain for the diagonal parts
\ba
&(D^t \rpm D_\Sigma)(f\rpm g)\rpm 2i \left [\V a_D \Delta \mp (\V b_D-\V a_D) f_3\right ]
\nonumber\\&
-2 i \left [\tilde D_{\V a}(\Sigma \Delta)\mp \tilde D_{\V b-\V a} (\Sigma f_3)\right ]
\mp (2 i)^2 \Delta {\Sigma \Sigma_z\over \Gamma} {\V a\times \V b}=0. 
\label{kin3}
\end{align}
The two equations for the off-diagonal Wigner functions $\Delta$ and $f_3$ become
\ba
f_3&={{i\over 2} D^t \Delta-\Sigma D_{\V a}f  -g\V a_D \over \Sigma-{\Sigma_z\over \Gamma} \V a_D}
\nonumber\\
\Delta&={{i\over 2} D^t f_3\mp \Sigma D_{\V b-\V a}f  \mp g(\V b_D-\V a_D) \over \Sigma-{\Sigma_z\over \Gamma} \V a_D}.
\label{off1}
\end{align}

\subsection{Influence of off-diagonal on diagonal elements}
In order to enclose the equations for the diagonal elements (\ref{kin3}) we have to find an approximate solution of the off-diagonal equations (\ref{off1}). We will consider only terms $o({\cal F})^2$ which means our theory will be valid up to quadratic order in derivatives or forces. One observes that $a_D^2\sim \V a_D D\sim D D\sim {\cal F}^2$ and we expand the denominator in (\ref{off1}) accordingly. In this way we can calculate the first required forms in (\ref{kin3})
\ba
&\Delta \V a_D \mp  f_3 (\V b_D-\V a_D)
=\mp [\V a_D \tilde D_{\V b-\V a}-(\V b_D-\V a_D) \tilde D_{\V a}] f
\nonumber\\&
=
\pm [{\cal F}\cdot \V b [\V a \p {\V r}+(\V a \times e\V B)]-
{\cal F}\cdot \V a [\V b \p {\V r}+(\V b \times e\V B)]]\p {\V k} f
\nonumber\\&
= \mp D_{{\cal F}\times (\V a\times \V b)}f
={i\over 2} D_{{\cal F}\times \V \Omega}f
\label{off2}
\end{align}
where we have used (\ref{rel1}) and (\ref{a10}).
Next we calculate $D^t\Delta, D^t f_3$ in order to introduce them mutually again in the expansion (\ref{off1}). The result reads
\ba
\Sigma \Delta&=
\mp g (\V b_D-\V a_D) 
\mp \Sigma \tilde D_{\V b-\V a}f+ o({\cal F}^2)
\nonumber\\
\mp \Sigma f_3&=
\pm g \V a_D  
\pm \Sigma \tilde D_{\V a}f + o({\cal F}^2).
\label{off3}
\end{align}
This allows to work out the second required form in (\ref{kin3})
\ba
\tilde D_{\V a}(\Sigma \Delta)&\mp \tilde D_{\V b-\V a}(\Sigma f_3)=
\mp g [\tilde D_{\V a}(\V b_D-\V a_D)-\tilde D_{\V b-\V a} \V a_D]
\nonumber\\&
\mp g [(\V b_D-\V a_D)\tilde D_{\V a} -\V a_D \tilde D_{\V b-\V a}]g
\nonumber\\&
\mp [\tilde D_{\V a}(\Sigma \tilde D_{\V b-\V a} f) -\tilde D_{\V b-\V a} (\Sigma \tilde D_{\V a} f)].
\label{form}
\end{align}
The first part of (\ref{form}) provides with (\ref{85})
\ba
&\mp g [\tilde D_{\V a}(\V b_D-\V a_D)-\tilde D_{\V b-\V a} \V a_D]\nonumber\\
&=
{i\over 2} g [\p {\V k} \times ({\cal F}\times \V \Omega)]\cdot e\V B
\pm g {\Sigma_z\over \Gamma} (\V b-\V a)\cdot {\cal F} (\V \Omega \cdot e\V B)
\end{align}
where the second part exactly cancels a part of the last term in (\ref{kin3})
\ba
&\pm (2 i) \Delta {\Sigma \Sigma_z\over \Gamma} {\V a\times \V b}\cdot e\V B 
\nonumber\\
&=\mp{\Sigma_z\over \Gamma} (\Omega \cdot e\V B)\left [{\cal F}\cdot  (\V b-\V a) g+\Sigma \tilde D_{\V b-\V a} f\right ].
\label{comp}
\end{align} 
The remaining part of (\ref{comp}) is canceled if we consider the third part of (\ref{form}) which takes with (\ref{83}) the form
\ba
&\mp [\tilde D_{\V a}(\Sigma\tilde D_{\V b-\V a} f) -\tilde D_{\V b-\V a} (\Sigma\tilde D_{\V a} f)]=
{i\over 2} \tilde D_{\V \Omega\times (e\V B\times \p {\V k} \Sigma)} f
\nonumber\\
&+{\Sigma \Sigma_z\over \Gamma}\tilde D_{\V b-\V a} f (\V \Omega\cdot e\V B)
\mp {i\over 2}\tilde D_{\Sigma \partial_k (\V \Omega\cdot e\V B)} f.
\end{align}
The second part of (\ref{form}) finally becomes with (\ref{a10})
\be
\mp g [(\V b_D-\V a_D)\tilde D_{\V a} -\V a_D \tilde D_{\V b-\V a}]g=\mp{i\over 2} \tilde D_{{\cal F}\times \Omega} g.
\ee
Collecting all pieces together we obtain for (\ref{kin3})
\ba
&(D^t\rpm \tilde D_\Sigma\rpm \tilde D_{{\cal F}\times \V \Omega}) (f\rpm g) 
\!+\!\tilde D_{\V \Omega\times (e\V B\times \p {\V k} \Sigma)} f
\!-\!\tilde D_{\Sigma \partial_k (\V \Omega\cdot e\V B)} f
\nonumber\\
&+g [\p {\V k} \times ({\cal F}\times \V \Omega)]\cdot e\V B=0.
\label{kin4}
\end{align}
As a check we see that all $\pm$ have dropped out and all Berry connections are condensed into a single Berry curvature which should be the case for physical quantities. In the following it is therefore sufficient to denote the equations for both diagonal Wigner functions as $\rpm=\pm$.

Now we will try to find an equation for the chiral Wigner functions itself. We have $f\pm g=f_{\pm\!\pm}$ such that we can write for both diagonal equations (\ref{kin4})
\be
(D^t\pm \tilde D+\tilde D_2) f_{\pm\!\pm}-(y\pm \tilde D_2)g=0
\label{ff2}
\ee
where it is convenient to use besides (\ref{D}) the abbreviations
\ba
\tilde D&=\tilde D_2+ \tilde D_{{\cal F}\times \V \Omega}
\nonumber\\
\tilde D_2&=\tilde D_{\V \Omega\times (e\V B\times \p {\V k} \Sigma)} f
\!-\!\tilde D_{\Sigma \partial_k (\V \Omega\cdot e\V B)}
\nonumber\\
F&=-[\p {\V k} \times ({\cal F}\times \V \Omega)]\cdot e\V B={\cal F}\cdot \p {\V k}(\V \Omega e B)={\cal F}\cdot \p {\V k} c. 
\label{abbrev}
\end{align}
We will abbreviate the repeatedly occurring product by $c=\V \Omega \cdot e\V B$.

\subsection{Disentanglement of diagonal Wigner functions}
The two equations (\ref{ff2}) we can use to express approximately $g$ by $f_{\pm\!\pm}$. Therefore we add both equations and use again $f=f_{\pm\!\pm}\mp g$ to obtain the iteration equation for $g$
\ba
g&=\frac 1 F (D^t+\tilde D_2)f_{\pm\!\pm}+\frac 1 F (\tilde D\mp D^t\mp \tilde D_2) g
\nonumber\\
&=\frac 1 F \left [1-(\tilde D\mp D^t\mp \tilde D_2) \frac 1 F\right ]^{-1} (D^t+\tilde D_2) f_{\pm\!\pm}
  \nonumber\\
  &={F\over F^2+(\tilde D\mp D^t\mp \tilde D_2) F} (D^t+\tilde D_2)f_{\pm\!\pm}+o(D^2)
    \nonumber\\
&={F\over (\tilde D\mp D^t\mp \tilde D_2) F} (D^t+\tilde D_2)f_{\pm\!\pm}  +o(F).
\label{gf}
\end{align}
Since $g$ appears in (\ref{ff2}) in front of forces it is allowed to express the last line in (\ref{gf}) in first-order forces $F$. Since we restrict to first-order gradients, we see that in the third line of (\ref{gf}) the operators act only on $F$. Introducing (\ref{gf}) into (\ref{ff2}) we obtain up to orders  $o(D^2, F)$
\ba
\left [D^t\pm \tilde D+\tilde D_2 \mp {\tilde D_2 F\over (\tilde D\mp D^t\mp \tilde D_2) F} (D^t+\tilde D_2\right ]f_{\pm\!\pm}&=0
\end{align}
where we neglect higher-order gradients as multiple applications of $D$s and $F^2$ terms.
In this sense we can also neglect products of $\tilde D_2 F \tilde D_2 f$ which leads finally to
\ba
\left [\left (1\mp{\tilde D_2 F\over \tilde D F}\right ) D^t\pm \tilde D+\tilde D_2\right ] f_{\pm\!\pm}=0
\label{ff4}
\end{align}
and we see that the approximate decoupling renormalizes the drift $D^t$. In fact remembering the abbreviations (\ref{abbrev}) we obtain for this factor up to higher-order gradients
\be
1\mp{\tilde D_2 F\over \tilde D F}&=&1\mp{-c (\p {\V k} \Sigma\times e\V B) \p {\V k} F-\Sigma (\p {\V k} c\times e\V B) \p {\V k} F \over (\p {\V k} \Sigma\times e\V B) \p {\V k} F}
\nonumber\\
&=&1\pm c+o(\p {\V k} c \p {\V k} F).
\label{off5}
\ee
Now we can rewrite the final kinetic equation (\ref{ff4}) introducing conveniently the renormalized quasiparticle energies
\be
\epsilon_\pm=\epsilon\pm \Sigma(1\mp c)
\label{eps}
\ee
which changes the drift (\ref{D}) as $D^t(\epsilon) \to D^t_\pm(\epsilon_\pm)$ and consequently
\be
{\cal F}_\pm={\cal F}\pm \p {\V k} [\Sigma(1\mp c)]\times e\V B
\ee
to obtain
\ba
\left \{
  (1\pm c)D^t_\pm \pm \tilde D_{{\cal F}_\pm \times\V \Omega}\pm [\V \Omega \p {\V k} (c\Sigma)]e\V B\cdot \p {\V r}
\right \}f_{\pm\!\pm}=0.
\label{ff5}
\end{align}

A short algebra translates now the final form (\ref{ff5}) into the phenomenological kinetic equation (\ref{phenom}) of the literature. Using $\V v=\p {\V k} \epsilon_\pm$ we rewrite the drift (\ref{D})
\ba
D^t_\pm \pm \tilde D_{{\cal F}_\pm \times\V \Omega}&=\p t+(\V v\pm (e\V E+\V v\times e\V B)\times \V \Omega)\cdot \p {\V r}
\nonumber\\
+&\left \{
  e\V E+[v\pm (e\V E+\V v\times e\V B)\times \V \Omega]\times e\V B\right \}\cdot \p {\V k}
\nonumber\\
&=\p t+D_\pm^M\mp c(\V v\cdot \p {\V r}+(e\V E+\V v\times e\V B)\cdot \p {\V k}
\nonumber\\
&=\p t+D_\pm^M\mp c(D_\pm^t-\p t))
\label{h1}
\end{align}
where we used the drift \cite{MT14}
\ba
D_\pm^M&=[\V v\pm e\V E\times \V \Omega \pm e\V B (\V v\cdot\V \Omega ]\cdot \p {\V r}
\nonumber\\
&+\left [
  e\V E+v\times e\V B \pm \V \Omega (e^2 \V E\times \V B)\right ]\cdot \p {\V k}.
\label{Dm}
\end{align}
From (\ref{h1}) one sees that
\be
(1\pm c) D_\pm^t\pm \tilde D_{{\cal F}_\pm \times\V \Omega}=(1\pm c) \p t +D_\pm^M
\ee
which we employ now in (\ref{ff5}). Dividing by $1\pm c$ we obtain the final kinetic equation
\ba
\biggl \{
  \p t &+ {1\over 1\pm c} D^M_\pm 
  \pm {1\over 1\pm c} [\V \Omega \p {\V k} (c\Sigma)]e\V B\cdot \p {\V r}
\biggr \}f_{\pm\!\pm}=0.
\label{ff6}
\end{align}
If we neglect the second line as higher order, we obtain exactly the phenomenological expected kinetic equation (\ref{phenom}) with the drift (\ref{rp}) which describes the trajectories (\ref{haldane}).

One can also rewrite $f_{\pm\!\pm}=(1\pm c) f_{\pm}$ in order to make the phase-space invariant  (\ref{G}) explicit. Before dividing by $(1\pm c)$, the kinetic equation (\ref{ff6}) translates then into 
\ba
&\biggl \{
 (1\pm c)  \p t + D^M_\pm 
\pm [\V \Omega \p {\V k} (c \Sigma)]e\V B\cdot \p {\V r}
\biggr \}f_{\pm}=
  \nonumber\\
  & f_\pm \left [{(e^2 \V E\cdot \V B)\V \Omega \cdot \p {\V k} c\!\pm\! {\cal F}\cdot \p {\V k} c \!+\!(1\!\mp\! c) (\p {\V k} \Sigma \!\times\! e\V B)\cdot \p {\V k} c \over 1\pm c}
    \right ].
\label{kin6}
  \end{align}
In the sense of the above approximations we can neglect the $\p {\V k} c=\p {\V k} (e\V B\cdot \V \Omega)$ terms on the right-hand side which go together with ${\cal F}$ or $\p {\V k}$ as higher order corrections. This means we obtain finally the chiral kinetic equation 
\ba
&\biggl \{
\p t + {1\over 1\pm e\V B \cdot \V \Omega}
\left [
    \V v\pm e\V E\times \V \Omega \pm e\V B (\V v\cdot\V \Omega)
    \right ]\cdot \p {\V r}
\nonumber\\
&+{1\over 1\pm e\V B \cdot \V \Omega}
  \left [
  e\V E+v\times e\V B \pm \V \Omega (e^2 \V E\times \V B)\right ]\cdot \p {\V k}
\biggr \}f_{\pm}=0
\label{kin7}
\end{align}
which is exactly (\ref{phenom}) with (\ref{rp}) and the normalization to the chiral densities
\be
n_\pm=\int {d\V k d \V r  \over (2\pi\hbar)^3}(1+e \V B \cdot \V \Omega) f_\pm.
\ee
Please note that we have neglected the additional spatial derivative term $[\V \Omega \p {\V k} (c \Sigma)]e\V B$ as being of higher order in (\ref{ff5}) or (\ref{kin6}).

\subsection{Summary of used approximations}
At this stage let us shortly recapitulate what kind of approximations had to be applied to derive the phenomenological equation (\ref{kin7}) from the spinor one (\ref{kin}). The starting equation (\ref{kin}) is approximated up to second order derivatives or gradients itself. 
In going from (\ref{off1}) to (\ref{off2}) and (\ref{off3}) we have used an expansion up to second order the in Lorentz force ${\cal F}$. The same approximation is employed when transforming (\ref{gf}) to (\ref{ff4}) together with the neglect of higher than second-order derivatives as used so far already. Equation (\ref{ff4}) takes the simple form (\ref{off5}) since we have considered a term $\p k (\Omega e B)\p k ({\cal F} \p k (\Omega e B))$ as being of higher order. 
In the same sense we finally have neglected the derivatives $\p k (\Omega e B)$ in (\ref{ff6}) and (\ref{kin6}). Here we restrict to Berry connections in momentum and neglect all spatial dependence due to the meanfield. This would lead to further spatial derivatives which could be worked out in the line presented here respecting also $\V a_r$ in $\V a_D$ (\ref{deriv}). 
The question is, however, whether the presented rewriting into Berry-curvature terms then is very sensible and whether it is not more convenient to work directly with the exact mean field equations (\ref{kin}) since they allow to include many-body effects systematically.  

One sees that the chiral kinetic equation (\ref{kin7}) or (\ref{phenom}) is by no means a simple rewriting of the meanfield kinetic equations (\ref{kin}). However, it leads to the same chiral anomaly. The difference is that the source of the chiral anomalous term is different in both equations. While from the chiral kinetic equation it comes exclusively from the zero momentum divergence of the Berry curvature and therefore Dirac monopoles, the spinor meanfield equations provide $1/3$ from the Dirac sea as we will see now.

\section{Chiral anomaly and magnetic monopoles}

\subsection{Chiral anomaly from exact meanfield kinetic equation}

The spinor kinetic equations (\ref{kin}) for SU(2) structures of chapter \ref{exact} can be used to describe Weyl systems. For this purpose we consider the limit of infinite mass of nonrelativstic particles which extinct the quasiparticle energy $\epsilon_k\to 0$ and only the chiral
Hamiltonian (\ref{ham}) remains with a proper choice of the vector selfenergy according to table~\ref{tab1}. This is the same procedure as was applied to describe graphene \cite{M16} and allows here to consider the kinetic equations for right and left-handed chiral particles $\V \Sigma =\pm v\V p$. Let us focus on the right-handed ones since the final result can be translated for left-handed particles by $v\to -v$. In order to see the chiral anomaly or equivalent forms, we will investigate the balance of the scalar density $n$ given by the momentum integral over the scalar Wigner function $f$.

We consider now the linear response of (\ref{kin}) with respect to an external potential $\phi$ and Fourier transform $\dot n(t,\V r) \to -i\omega n(\omega,\V q/\hbar)$. The linearized coupled equations (\ref{kin}) read
\ba
-i\omega\hbar \delta \V g\!-\!ie\phi \V q \p {\V k} \V g_0\!+\!e v(\V B\!\times\! \p {\V k})f_0\!+\!i v \V q \delta \V g&=
2 (\V \Sigma\!\times\! \delta \V g)
\nonumber\\
-i\omega\hbar \delta f\!-\!ie\phi \V q \p {\V k} f_0\!+\!e v(\V B\!\times\! \p {\V k})\V g_0\!+\!i v \V q \delta \V g&=0
\label{dfg}
\end{align}
where we have used
\ba
\delta f&=\delta f_0-{e\over \hbar\omega} \phi \V q \p {\V k} \delta f_0-{ie v\over \hbar\omega}\V B\times \p {\V k} \delta \V g_0
\nonumber\\
\delta \V g&=\delta \V g_0-{e\over \hbar\omega} \phi \V q \p {\V k} \delta \V g_0-{ie v\over \hbar\omega}\V B\times \p {\V k} \delta f_0.
\label{fg}\end{align}
Here the subindex $\delta f_0$ denotes the linearization terms without electric and magnetic fields. In this way we concentrate only on the linear electric field response since we are searching for $\V E\cdot \V B$ terms. The complete response functions are given in \cite{M15a}. The advantage of writing the linearization in the two-step form (\ref{fg}) and (\ref{dfg}) is that the scalar density response becomes
\be
-i\omega \delta n_+=-{i\over \hbar}\V q \delta \V j_{+}-e v\int {d^3k\over (2\pi\hbar)^3} \V B\times \p {\V k} \delta \V g_0
\label{dn}
\ee
where $-\omega \sum_k \delta f_0=-\omega \delta n_0=\V q \delta \V j_{+}$ and we have to calculate $\delta \V g_0$ from (\ref{dfg}) only in zeroth order of the magnetic field. This latter vector equation (\ref{dfg}) is readily solved e.g. with the help of the formula (A3) of \cite{M15a} to yield
\ba
&\delta \V g={-{e\over \omega\hbar }\over 1-{4 v^2k^2\over \hbar^2\omega^2}}\biggl \{
\phi \V q\!\cdot\! \p {\V k} \V g_0
\!+\!{2 i v k\over \hbar \omega}\phi  \V e \!\times\! \V q \!\cdot\! \p {\V k} \V g_0
\nonumber\\
&-{4v^2k^2\over \hbar^2\omega^2} \phi \V e (\V e \V q\!\cdot\! \p {\V k} \V g_0)
\!+\!i v (\V B\!\times\! \p {\V k}) f_0
\nonumber\\
&  -{2 v^2 k\over \hbar\omega}\V e \!\times\! (\V B\!\times\! \p {\V k} f_0)
-{4iv^3k^2\over \hbar^2\omega^2} \V e [\V e \!\cdot\! (\V B \!\times\! \p {\V k} f_0)]
\biggr \}
\label{dg}
\end{align}
where we use $\V e=\V \Sigma/\Sigma=\V k/k$.
We introduce now (\ref{dg}) into (\ref{dn}) and restrict to linear terms in the magnetic field.
Observing that $\V g_0=\V e g_0$ and $\V q\p {\V k} \V g_0=g_0q\p {\V k} \V e+\V e q \p {\V k} g_0$ as well as $\V q\p {\V k} \V e=(\V q-\V e (\V q \V e))/k$ we obtain for (\ref{dn})
\ba
\delta n_+=&\delta n_{0+}+{i e^2 v \over \hbar\omega^2 }\phi \int {d^3k\over (2\pi\hbar)^3} \V B\times \p {\V k}
\biggl\{
\V e \V q\cdot \p {\V k} g_0
\nonumber\\&
+{g_0\over k\!-\!{4 v^2k^3\over \hbar^2\omega^2}} \left [
  \V q\!-\!\V e (\V q\V e)\!+\!2 i {v k\over \hbar\omega}(\V e \!\times\! \V q)\right ]
\biggr \}.
\label{dnn0}
\end{align}
The volume integral is transformed into a surface integral with the help of a modified Gau\ss{}-Ostrogradsky relation
\be
\int d^3k (\V B\times \p {\V k})\V \Psi=\V B\cdot \int d\V A\times\V \Psi.
\ee
The surface element $d\V A=\V e k^2 \sin{\theta} d\theta d\varphi$ denotes a sphere with radius $k$.
We obtain for (\ref{dnn0})
\be
-i \omega \delta n_+ + {i\over \hbar}\V q \delta \V j_{+}=R(\infty)-R(0)
\label{dnn}
\ee
where we abbreviate the surface integral with radius $k$
\ba
R(k)\!=\!
{i e^2 k^2\phi\V B\over {\hbar^2\omega^2\over 4 v^2}\!-\!k^2}\!\int\limits_0^{2\pi}\!\!\! {d\varphi\over (2\pi\hbar)^3}\!\!\int\limits_{-1}^1\!\!\!d x g_0 \!\left \{\!
    {\V e\!\times \!\V q\over v k}\!+\!{2 i\over \omega}[\V e (\V e \V q)\!-\!\V q]\!\right \}
  \end{align}
  and $x=\cos{\theta}$. We consider $\V q$ as the $z$-axes of integration such that the $d\varphi$ integration renders the term $\V e \times \V q$ zero and we obtain with $\V E=-i \V q \phi/\hbar$
\be
R(k)=\frac 1 4 R_c{k^2\over k^2-{\hbar^2\omega^2\over 4 v^2}}\int\limits_{-1}^1 d x (x^2-1)\, g_0.
  \ee
The term of the chiral anomaly (\ref{rate}) we introduce as
\be
R_c= {e^2\over 2 \pi^2\hbar^2} \V E\V B.
\ee
Now we are ready to see the different sources of the $\V E\V B$ term. We interpret the $f_-$ Wigner function in (\ref{solF}) as a hole or antiparticle
\be
f_-=1-\bar f_+
\ee
with
\be
\bar f_+(\epsilon_+)={1\over {\rm e}^{{\epsilon_++\mu\over T}+1}}
\ee
instead of particle Wigner function $f_+(\epsilon_+)$ where $\mu \to -\mu$. In such a way
\be
g_0=\frac 1 2 (f_+-f_-)=\frac 1 2 (f_++\bar f_+-1).
\label{sea}
  \ee
We can now evaluate the limits in (\ref{dnn}). 

First we consider the simple dispersion $\epsilon_+=v k$. Then the $x$-integration is trivial and one gets
\be
R(\infty)=\frac 1 6 R_c,\qquad R(0)=0
\ee
and we see that the anomalous term comes exclusively from the $k\to \infty$ limit which is the $-1$ term in (\ref{sea}). Therefore it comes from the Dirac sea and not from the Dirac monopole at $k\to 0$.

The situation is changed if we consider the modified dispersion resulting from the rewriting of the kinetic equation in form of chiral equation of chapter (\ref{transform}). This means we have from (\ref{eps})
\be
\epsilon_+=v k-v k e\V B\cdot \V \Omega=v k-{e v\hbar \V B\cdot \V k\over 2k^2}=v k-{e v \hbar B x\over 2k}.
\label{eps1}
\ee
Then the $x$ integration is more involved but still analytical (appendix~\ref{integral}) and we obtain a non-vanishing limit at $k\to 0$
\be
R(0)=\left \{
  \begin{array}{rcl} -\frac 1 3 R_c&{\rm for} & \omega=0 \cr 0 &{\rm for} & \omega\ne 0\end{array}
  \right .
  \ee
only for the long-time or static case.
This means for the generalized dispersion with magnetic moment interaction by the Berry curvature $\Omega$ we obtain the chiral anomaly in the balance equation for the right-handed density (\ref{dnn})
\be
-i\omega \delta n_++ {i\over \hbar}\V q \delta \V j_{0+}=\left \{
  \begin{array}{rcl} \frac 1 2 R_c&{\rm for} & \omega=0 \cr \frac 1 6 R_c &{\rm for} & \omega\ne 0\end{array}
  \right .
  \ee

  The kinetic equation for the left-handed chiral particles are identical just by replacing $v\to -v$ which results into $g_0\to -g_0$. Therefore the difference of the chiral right- and left-handed densities becomes
\ba
-i\omega \delta (n_+-n_-)+ {i\over \hbar}\V q (\delta \V j_{+}-\V j_{-})=\left \{
  \begin{array}{rcl} R_c&{\rm for} & \omega=0 \cr \frac 1 3 R_c &{\rm for} & \omega\ne 0\end{array}
  \right ..
  \end{align}
  This is a remarkable result by two aspects. First we see that the static or long time-limit is different from the dynamical result. While the static limit agrees with the chiral anomaly reported in the literature, the finite frequency or dynamics leads to $1/3$  which would mean a topological charge (\ref{topo}) of $1/3$.
Recently there has been found an additional dynamical part arising from the magnetization current as the curl of the magnetization \cite{KSY17} which exactly fills the missing part of the dynamical result. The second aspect is that the anomalous term does not originate exclusively from the vanishing momentum and Dirac monopole but $1/3$ comes from the Dirac sea which is the $k\to \infty$ limit.

  \subsection{Chiral anomaly from chiral kinetic equation}

The density balance of the chiral kinetic equation (\ref{kin7}) are exactly leading to (\ref{balancerate}) with \cite{MT14}
\be
\dot n_\pm+\nabla (\V j_\pm+\V E\times \V \sigma_s)=\xi{e^2\over 4 \pi^2\hbar^2} \V E\cdot \V B.
\ee
The same result appears from the antiparticles where $\mu\to -\mu$ such that we can add a factor 2 on the right-hand side and since $\p {\V k}\V \Omega=0$ we have for the Chern number
\ba
&\xi
= \pm 2 \!\!\int\!\!  {d^3 k\over 2 \pi} \V \Omega \p {\V k} f_+
=\pm\!\!\int\!\! {d^3 k\over \pi} \p {\V k} ( \V \Omega f_+)
=\pm \!\!\int\limits_{k=0}^{k=\infty}\!\! {d \V A\over \pi}\V \Omega f_+
\nonumber\\
&=\pm \left .   \!\!\int\limits_{-1}^1\!\! {d x \over {\rm e}^{{v k\over T}-{e v \hbar B x\over 2k T}-{\mu\over T}}\!+\!1}\right |_{k=0}^{\infty}
\!\!\!\!\!=\left .
  \pm {2 T k\over e v B} \ln{
    {
      {\rm e}^{e v \hbar B \over 2k T}\!+\!{\rm e}^{{v k\over T}\!-\!{\mu\over T}}\over
      {\rm e}^{-e v \hbar B \over 2k T}\!+\!{\rm e}^{{v k\over T}\!-\!{\mu\over T}}
    }
    }
  \right |_{0}^{\infty}
  \nonumber\\
&
\to \pm 1
\end{align}
where we have used (\ref{eps1}) and the $k\to\infty$ limit is zero. Therefore the chiral anomaly term comes exclusively from the zero momentum or Dirac monopole in the chiral kinetic theory.

\section{Summary and conclusions}

We have considered the approximating steps to derive the chiral kinetic equation from the exact spinor mean field kinetic equation. It turns out that the Berry connection can be given in terms of the vector self energy which comprises the meanfield, the magnetic field and the spin-orbit coupling vector.
The chiral density balance shows a seemingly non-conservation which source is identified to originate with $1/3$ from the Dirac sea at infinite momentum and with $2/3$ from the Berry curvature at zero momentum which would be the Dirac monopoles. The origin by the Dirac sea is transferred to the Dirac monopole during the rewriting in chiral basis and the resulting chiral kinetic equation. The dynamical result suppresses $2/3$ of the chiral anomaly compared to the static or long-time limit which can be compensated by extra currents from magnetization. Interestingly this chiral anomalous $\V E\V B$ term is independent of temperature and density since we have derived the kinetic theory for finite temperatures and densities.

We obtain the same chiral anomalous terms from a conserving chiral Hamiltonian as it appears by Adler-Jackew-Bell or triangle anomaly or by the assumption of symmetry-breaking axion fields in the Lagrangian. We show here that no symmetry-breaking assumptions are necessary to produce such $\V E\V B$ term violating chiral density balance. In other words the experimental verification of such term does not allow to conclude on Lorentz-symmetry breaking or mixed chiral-gravitational anomaly.

The linear response terms of the spinor kinetic equation can be alternatively used to describe the experiments as it was possible from chiral kinetic equation since their equivalence are shown here and the necessary approximations. Deviations of the transport results, once calculating by spinor kinetic equation and once by the chiral kinetic equation, should now be worked out further in detail but exceeded the size of this paper.

As a note in the end, the chiral kinetic equation can be written in terms of covariant derivatives and the Berry-connection as arising from the overlap of Bloch wave functions \cite{Ha93}. For an overview see \cite{CN06}. The treatment here is therefore equivalently applicable to Bloch electrons in semiconductors.

\appendix
\section{Calculation of derivatives}

The derivatives (\ref{deriv}) are needed for momentum-dependent $\V a$ and $\V b$. A straight calculation with (\ref{deriv}) shows the form
\ba
\V b_d\tilde D_\V a\!-\!\V a_d \tilde D_\V b\!=\!\tilde D_{(\V b \cdot {\cal F}) \V a\!-\!(\V a\cdot {\cal F}) \V b}=\tilde D_{{\cal F} \!\times\! (\V a\!\times \!\V b)}=
-\frac i 2 \tilde D_{{\cal F} \!\times\! \V \Omega}
\label{a10}
\end{align}
where
(\ref{rel1}) has been used.

Since only first-order gradients are considered, the occurring multiple operations reads
\ba
\tilde D_{\V a}\da {\tilde D_{\V c}}f&=[\V a \cdot \p {\V r}+(\V a \times e\V B)\cdot \p {\V k}] [\da{\V c} \cdot \p {\V r}+(\da {\V c} \times e\V B)\cdot \p {\V k}] f
\nonumber\\
&=(\V a\times e\V B)\cdot \p {\V k} (\da {\V c}\cdot \V d)
\end{align}
where we have used a simple rotation of the scalar triple product and introduce
\be
\V d=\p {\V r} f+e\V B\times \p {\V k} f.
\label{d}
\ee
We can proceed
\ba
&\tilde D_{\V a}\da {\tilde D_{\V c}}f=
(\V a\times e\V B)\cdot [\V d\times (\p {\V k} \times {\V c})+(\V d\cdot \p {\V k}) \V c]
\nonumber\\
&=\V d\cdot \bigl \{(\p {\V k}\times \V c) \times (\V a\times e\V B)+ \p {\V k}[(\V a \times e\V B)\cdot \da {\V c}\bigr \}
\nonumber\\
&=\V d\cdot \bigl \{(\V a \times e\V B)\cdot \p {\V k} {\V c}\bigr \}.
\end{align}

Similarly we calculate after rotation of the triple scalar product and introducing (\ref{d})
\ba
&(\tilde D_{\V a}\tilde D_{\V b}-\tilde D_{\V b}\tilde D_{\V a})f=(\V a \times e\V B)\cdot \p {\V k} (\da {\V b}\cdot \V d)-(a\leftrightarrow b)
\nonumber\\
&=[\p {\V k} (\da {\V b}\cdot \V d)\times \V c]\cdot e\V B-(a\leftrightarrow b)
\nonumber\\
&=\bigl\{
\p {\V k}\times [(\da{\V b}\V d) \da {\V a}]-(\V b \V d) \p {\V k}\times \V a-(a\leftrightarrow b)\bigr \} \cdot e\V B
\nonumber\\
&=\bigl\{
\p {\V k}\!\times \! [\V d \!\times\! (\da{\V a}\!\times\! \da{\V b})]\!-\! (\V b \cdot\V d) \p {\V k}\!\times\! \V a \!-\! (\V a \cdot\V d) \p {\V k}\!\times\! \V b \bigr \} \cdot e\V B
\label{80}
\end{align}
Now we can use (\ref{rel1}) to obtain
\ba
&(\tilde D_{\V a}\tilde D_{\V b}-\tilde D_{\V b}\tilde D_{\V a})f
\nonumber\\
&=-\frac i 2 [\p {\V k} \times (\V d \times \da{\V \Omega})]\cdot e\V B
\mp {\Sigma _z\over \Gamma} (\V b-\V a) \cdot \V d (\V \Omega \cdot e\V B)
\nonumber\\
&
=\frac i 2 (\V d\cdot\p {\V k}) (\V \Omega\cdot e\V B)
\mp {\Sigma _z\over \Gamma} (\V b-\V a) \cdot \V d (\V \Omega \cdot e\V B)
\nonumber\\
&
=\frac i 2 \tilde D_{\p {\V k} (e\V B\V \Omega)} f\mp {\Sigma _z\over \Gamma} \tilde D_{\V b-\V a} f.
\label{82}
\end{align}

Another occurring expression is
\ba
&\tilde D_{\V b} f\tilde D_{\V a} c-\tilde D_{\V b} f\tilde D_{\V a} c
\nonumber\\
&=
[\p {\V r} f\!\times \! (\V a\!\times \!\V b)\!+\!(e\V B\times \p {\V k} f)\!\times\! (\V a\!\times\! \V b)](\p {\V r} c\!+\!e\V B\times \p {\V k} c)
\nonumber\\
&=-\frac i 2 [\V \Omega \times (\p {\V r} c+e\V B\times \p {\V k} c)](\p {\V r} f+e\V B\times \p {\V k} f)
\nonumber\\
&=-\frac i 2 \tilde D_{\V \Omega \times (\p {\V r} c+ e\V B\times \p {\V k} c)} f.
\label{81}
\end{align}

Finally one needs
\ba
&\tilde D_{\V a} (c \tilde D_{\V b} f)-\tilde D_{\V b} (c \tilde D_{\V a} f)
\nonumber\\
&=(\tilde D_{\V a} \da{c}) \tilde D_{\V b} f+c \tilde D_{\V a} \da {\tilde D_{\V b}} f-(a\leftrightarrow b)
\nonumber\\
&=-\frac i 2 \tilde D_{\V \Omega \times (\p {\V r} c+ e\V B\times \p {\V k} c)} f+\frac i 2 \tilde D_{\p {\V k} (e\V B \V \Omega)}f
\nonumber\\
&\quad \mp {\Sigma _z\over \Gamma} (e\V B\V \Omega) c\tilde D_{\V b-\V a} f
\label{83}
\end{align}
where we used (\ref{81}) and (\ref{82}).

The expression with (\ref{deriv})
\be
\tilde D_{\V a} \V b_D-\tilde D_{\V b} \V a_D=(\V a\times e\V B)\p {\V k} (\V b\cdot {\cal F})-(a\leftrightarrow b)
\ee 
leads to a form (\ref{80}) with $\V D={\cal F}$ such that we obtain
\ba
\tilde D_{\V a} \V b_D-\tilde D_{\V b} \V a_D&=
-\frac i 2 [\p {\V k} \times ({\cal F} \times \da{\V \Omega})]\cdot e\V B
\nonumber\\
&\mp {\Sigma _z\over \Gamma} (\V b-\V a) \cdot {\cal F} (\V \Omega \cdot e\V B).
\label{85}
\end{align}

\section{Angular integration}\label{integral}

The angular integration
\ba
I=\int\limits_{-1}^1 d x (x^2-1) [{\rm e}^{{v k-\mu\over T} +a x}  +1]^{-1}
\label{b1}
\end{align}
can be performed
where we abbreviated
\be
a={e v B\over 2 k T}.
\ee
Therefore the $x^2$ term in (\ref{b1}) is represented by a second derivative of
\be
{\partial^2\over \partial a^2}\left [
  a{k v x\over T}-Li_2(-{\rm e}^{{m-k v\over T}+a })
  \right ]={x^2\over {\rm e}^{{v k-\mu\over T} +a x}  +1}
\ee
with respect to $a$ which we will perform after the $x$-integration. Here the polylogarithm function $Li_n(x)=\sum_{k=1}^{\infty} z^k/k^n$ is used. The $-1$ term in (\ref{b1}) is trivial. Performing the integration and derivations we obtain
with the abbreviation $b=(k v-\mu)/T$
\be
I=&&{2\over a^3} \left \{
  a\left [
    Li_2(-{\rm e}^{a\!-\!b})+Li_2(-{\rm e}^{a\!+\!b})
  \right ]
\right .\nonumber\\
&&\left .
 -Li_3(-{\rm e}^{a\!-\!b})+Li_3(-{\rm e}^{a\!+\!b})
\right \}
\nonumber\\
\to&& -\frac 2 3\, {\rm for}\,k\to 0
\ee
where the $k\to 0$ limit appears by the $b\ll a$ limit.

\bibliography{entropy,bose,kmsr,kmsr1,kmsr2,kmsr3,kmsr4,kmsr5,kmsr6,kmsr7,delay2,spin,spin1,refer,delay3,gdr,chaos,sem3,sem1,sem2,short,cauchy,genn,paradox,deform,shuttling,blase,spinhall,spincurrent,tdgl,pattern,zitter,graphene,quench,msc_nodouble,iso,march,weyl}
\bibliographystyle{../../script/apsrev_meins}

\end{document}